\documentstyle[epsf,amsfonts,amssymb]{europhys}

\newif\ifboo \boofalse

\begin{document}
\euro{}{}{}{}
\Date{Version of April 7, 1999}
\title{Energy Landscape and Overlap Distribution of Binary
 Lennard-Jones Glasses}
\shorttitle{Energy Landscape of LJ Glasses}
\author{K. K. Bhattacharya, K. Broderix, R. Kree and A. Zippelius}
\institute{Institut f\"ur Theoretische Physik,
           Universit\"at G\"ottingen,\\
           Bunsenstr.\ 9,
           D-37073 G\"ottingen,
           Germany}
\rec{}{}
\pacs{\Pacs{61}{43.Fs}{Glasses}\Pacs{64}{70.Pf}{Glass transitions}}
\maketitle
\begin{abstract}%
  We study the distribution of overlaps of glassy minima, taking proper care of
  residual symmetries of the system.  Ensembles of locally stable, low lying
  glassy states are efficiently generated by rapid cooling from the liquid
  phase which has been equilibrated at a temperature $T_{\rm run}$.  Varying
  $T_{\rm run}$, we observe a transition from a regime where a broad range of
  states are sampled to a regime where the system is almost always trapped in a
  metastable glassy state. We do not observe any structure in the distribution
  of overlaps of glassy minima, but find only very weak correlations,
  comparable in size to those of two liquid configurations.
\end{abstract}

The phenomenology of supercooled liquids, the glass transition and the
glassy phase have been related to results of analytical calculations
for microscopic models of spin glasses \cite{kt87,p97}. The latter
bear no obvious relevance to structural glasses, hence, it is
particularly important to see, whether the analogy can be supported
{\it quantitatively}. One of the most prominent features of those
spin-glass models, which have been put forward as models of structural
glasses, is one step replica symmetry breaking, resulting in a bimodal
distribution of overlaps between pure states. Generally speaking, one
would expect to see non-trivial correlations among glassy states, if
mean field theory of spin glasses were to apply.

The liquid to glass transition as it is observed {\it e.g.} in metallic
glasses, is a non-equilibrium phenomenon with the crystalline state being the
true ground state.  Glassy states are in general metastable states, sometimes
called quasi-ergodic components, which are mutually inaccessible on
experimental time scales, while equilibrium within a quasi-ergodic component is
reached on much shorter time scales. In contrast to mean field models of spin
glasses the ensemble of metastable states depends on the preparation method and
the appropriate weights for metastable states are not unique. The assumption of
an idealised preparation method, which cools the liquid arbitrarily slowly and
at the same time prevents crystallization, has never been verified and seems
contradictory to us. We therefore prefer to use a different idealised
preparation ensemble which corresponds to infinitely rapid cooling from liquid
equilibrium states. This method was first introduced by Stillinger and Weber
\cite{sw82,sw84}.

\section{Distance Measure in Configuration Space}
We study correlations among an ensemble of such metastable states,
labelled by $a$.  A glassy state $a$ is characterized by a spontaneous
breaking of translational invariance as signalled by a non-zero
expectation value of the Fourier-components
$\langle\exp{(i\vec{k}\vec{r}_{i})}\rangle_a\neq 0$ of the local
density. Here $\vec{r}_i$ denotes the position of the $i$-th particle,
$i=1,\dots,N$. The local density is analogous to the local
magnetization, which is non-zero in the spin glass phase, $\langle
s_i\rangle_a\neq 0$, indicating the spontaneous breaking of the Ising
symmetry. An overlap between two glassy configurations can be defined
as
\begin{equation} 
  Q^{a,b}(\vec{k}) = 
  \frac{1}{N}\sum_{i=1}^{N}\langle\exp{(i\vec{k}\vec{r}_{i})}\rangle_a
  \,\langle\exp{(-i\vec{k}\vec{r}_{i})}\rangle_b
\end{equation}
in naive analogy to the spin glass order parameter
$Q^{a,b}=\frac{1}{N}\sum_{i=1}^{N}\langle s_i\rangle_a\langle
s_i\rangle_b$. For the glassy state, however, $Q^{a,b}(\vec{k})$ is not a good
measure, because it depends on the labelling of the particles. In particular
two identical configurations with different labels can have a very small
overlap according to the above definition. Permutations of identical particles
do not give rise to new states in the space of physically distinguishable
configurations. Hence, we need a distance measure which is invariant under all
permutations $\pi$ of identical particles in one state only.  This can be
achieved by maximizing over all such permutations
\begin{equation} 
  Q^{a,b}(\vec{k})=\max_{\pi}\frac{1}{N}\sum_{i=1}^{N}
  \langle\exp{(i\vec{k}\vec{r}_{i})}\rangle_a
  \,\langle\exp{(-i\vec{k}\vec{r}_{\pi(i)})}\rangle_b.
\end{equation}
This overlap has the required invariance property, but its computation requires
the enumeration of all permutations, which is computationally hard and can at
most be done for very small systems.

The computationally hard problem can be avoided by working with densities. We
specialize to zero temperature and only consider configurations of locally
minimal energy. Such a configuration is uniquely characterized by a point
measure $\rho^a(\vec{r})=(1/N)\sum_{i=1}^{N}\delta(\vec{r}-\vec{r}^{\,a}_{i})$,
which by construction is independent of the labelling of the particles. To
obtain a distance measure in configuration space, we regularise the point
particle densities $\delta(\vec{r})$ to homogeneous $\eta$-spheres
\begin{equation} 
  \delta(\vec{r}) \rightarrow \Delta_{\eta}(\vec{r}) = 
  \frac{1}{V_{\eta}} \theta(\eta-|\vec{r}|),
\end{equation} 
Here $V_{\eta}=4\pi\eta^3/3$ is the volume of the $\eta$-sphere and
we choose $2\eta<r_{\rm min}^a
=\min_{i,j}(|\vec{r}^{\,a}_{i}-\vec{r}^{\,a}_{j}|)$ to guarantee that the
spheres do not overlap and that the positions $\vec{r}^{\,a}_{i}$ can still be
uniquely reconstructed from the regularised density. Such a regularisation is
necessary because products of $\delta$ functions are ill defined and the
probability of coincidence of any two 3-dimensional vectors which are
distributed smoothly in space is equal to zero.

A natural distance measure between two regularised densities is
\begin{equation}
  Q^{a,b}= \frac{V_{\eta}}{N}
  \int {\rm d}^3x \sum_{i,j=1}^{N}\Delta_{\eta}
  (\vec{x}-\vec{x}^{\,a}_i) \Delta_{\eta}
  (\vec{x}-\vec{x}^{\,b}_{j}).
\end{equation}
If we restrict $\eta$ to even smaller values, $4\eta<r_{\rm
min}:=\min(r_{\rm min}^a, r_{\rm min}^b)$, then each sphere in state $a$ can
overlap with at most one sphere in state $b$.  Consequently the product of two 
local regularised densities is only non-zero, for the permutation $j=\pi(i)$,
which identifies particles within the same $\eta$-sphere.  The overlap is
then given by the volume fraction of overlapping spheres in the two
configurations, so that $0\leq Q^{a,b}\leq 1$.

It is obvious that $\eta$ cannot be made arbitrarily small, because then
the distance measure looses its ability to discriminate between different
structures. Very small $\eta$ will simply lead to maximal distances in most
cases and ultimately reproduce the point measure. We have varied $\eta$
over the range $0<\eta<r_{\rm min}/4$ and computed the number of
overlapping spheres for two configurations which are random and two
configurations which are similar (about 80\% overlap). We find that the number
of overlapping spheres is hardly sensitive to the choice of $\eta$ in the
range $r_{\rm min}/8<\eta<r_{\rm min}/4$.  For $\eta<r_{\rm min}/8$ the
number of overlapping spheres decreases drastically with $\eta$ even for
the strongly correlated configurations so that $Q^{a,b}$ can no longer
discriminate between strongly and poorly correlated states. In the following we
fix $\eta=r_{\rm min}/5$.

Other distance measures have been used in the literature \cite{heuer,cp98}, in
particular
\begin{equation}\label{mindist}
  D(a,b)= \min_{\pi} \sum_{i=1}^N 
  \left(\vec{r}^{\,a}_{i}-\vec{r}^{\,b}_{\pi(i)} \right)^2 , 
\end{equation} 
which requires the solution of a hard computational problem. It has the further
disadvantage to be not directly related to quantities which may be obtained
from experiment (like the density) or which appear naturally in any of the
existing theories of supercooled liquids and structural glasses.

A system in free space with pairwise interactions possesses translational and
rotational symmetries, which may be broken explicitly by boundary
conditions. The most convenient boundary conditions for computer simulations,
namely a cubic box with periodic boundary conditions, leave a subgroup of the
free space symmetries unbroken. This subgroup $\cal{S}$ is generated from the
translations and the cubic point group symmetries. As a consequence, there
exists a whole orbit of symmetry related, degenerate minima of $H$ to every
single minimum found in the simulations.

We are interested in the rate of occurrence, {\it i.e.} the probability
distribution, of overlaps $P(Q)=\sum_{a,b}P_aP_b\,\delta(Q-Q^{a,b})$.  Here
the summation includes all symmetry related states, which all have the {\it
same rate of occurrence} $P_a$. The large number of symmetry related states
imposes two severe difficulties.
First, many translated states (out of a continuum of states) have to be
generated to obtain a reliable estimate of $P(Q)$ from
numerical simulations.  Second, the $Q$ values
generated by all symmetry related states may scatter considerably in the
interval $0\leq Q\leq 1$. This may smear out structures in $P(Q)$ which would
be clearly marked in an ensemble without residual symmetries. To avoid this
problem we have broken the translational symmetry explicitly by fixing the
centre-of-mass. Since total momentum is conserved in our simulations, this can
be achieved by an appropriate choice of initial conditions (see below). The
cubic point group symmetry has been left untouched: for each minimum which was
found by our algorithm, we have generated 48 equivalent states and included
them in the histogram for $Q^{a,b}$. To check, whether this procedure smears
out relevant structure, we have also maximized the overlap over all 48
equivalent states.

\section{Simulations}
The system under consideration is a binary mixture of large (L) and
small (S) particles with $80\%$ large and $20\%$ small particles.
Small and large particles only differ in diameter, but have the same
mass. They interact via a Lennard--Jones potential of the form
$U_{\alpha\beta}(r)=4\,\epsilon_{\alpha\beta}[(\sigma_{\alpha\beta}/r)^{12}
- (\sigma_{\alpha\beta}/r)^{6}]$. All results are given in reduced
units, where $\sigma_{LL}$ was used as the length unit and
$\epsilon_{LL}$ as the energy unit. The other values of $\epsilon$ and
$\sigma$ were chosen as follows: $\epsilon_{LS} = 1.5, \sigma_{LS} =
0.8, \epsilon_{SS} = 0.5, \sigma_{SS} = 0.88$. The systems were kept
at a fixed density $\rho\approx 1.2$. Periodic boundary conditions
have been applied and the potential has been truncated appropriately
according to the minimum image rule \cite{at86}, and shifted to zero
at the respective cutoff.  For the systems with $N=30$ particles the
cut-off is $r_c = 1.43$ and for $N=60$ the cut-off is at $r_c = 1.7$.
The choice of the Lennard-Jones parameters follows recent simulations
of Lennard-Jones glasses \cite{ka94,vkb96}; it is known to suppress
recrystallization of the system on molecular dynamics time scales. The
glass transition is supposed to occur at the temperature $T_g\approx
0.45$ \cite{ka94}.  Throughout this study we present results for
systems with $N=30$ and $N=60$ particles, noticing that most of the
results have been verified for $N=100$.

Initially $N$ atoms are placed randomly inside a cubic simulation box of
side-length $L$.  The following steps are performed repeatedly:
\begin{itemize}
\item[1] Heat up the system to a temperature $T_{\rm run}$.
\item[2] Let the system evolve for a time $\tau_{\rm run}$ using
  molecular dynamics.
\item[3] Locate the nearest local minimum by quenching down the system
  to $T=0K$ using the steepest descent path.
\end{itemize}

Explicit breaking of translational invariance can be achieved by fixing the
centre-of-mass of the system. 
A unique definition of the centre-of-mass $\vec{R}_s$ is not obvious for
periodic boundary conditions. We start from the extended zone scheme
representation of the simulation box. The infinite periodic density pattern is
cut off at finite volume $V$, so that the cubic point group symmetry stays
intact.  The centre-of-mass is the same for all such volumes, including the
smallest one, the simulation box itself. The regularised $\vec{R}_s$ remains
unchanged in the limit $V\to \infty$.

Each run is initialized with the {\it same} positions and different
velocities of the particles. The latter are drawn from a
Maxwell-Boltzmann distribution with temperature $T_{\rm run}$, subject
to the constraint that the total momentum vanishes. The total momentum
is conserved in the molecular dynamics simulation as well as in the
steepest descent algorithm. Thereby we generate an ensemble, such that
in all states the centre-of-mass stays fixed. In the following we
shall also consider states which are related by cubic point group
symmetries. To guarantee that the centre of mass remains unaffected by
point group operations we choose $\vec{R}_s=0$.  We use standard
molecular dynamics with the velocity form of the Verlet algorithm
\cite{at86}. Steepest descent is achieved with the conjugate gradient
algorithm \cite{nr}.

\section{Results}
Simulating small systems requires careful checks to avoid sampling crystalline
configurations. First, it can be quite helpful to look at the
structures. Second, we have applied a common neighbor analysis \cite{ha87} to
detect the amount of polytetrahedral order (amorphous) and 
closed-packed structures (crystalline). 
Occasionally the structural stability of low energy states has been verified
for temperatures below the glass transition.

\begin{figure}[htbp]
  \begin{center}
    \leavevmode 
    \epsfysize=5.5cm\epsfbox{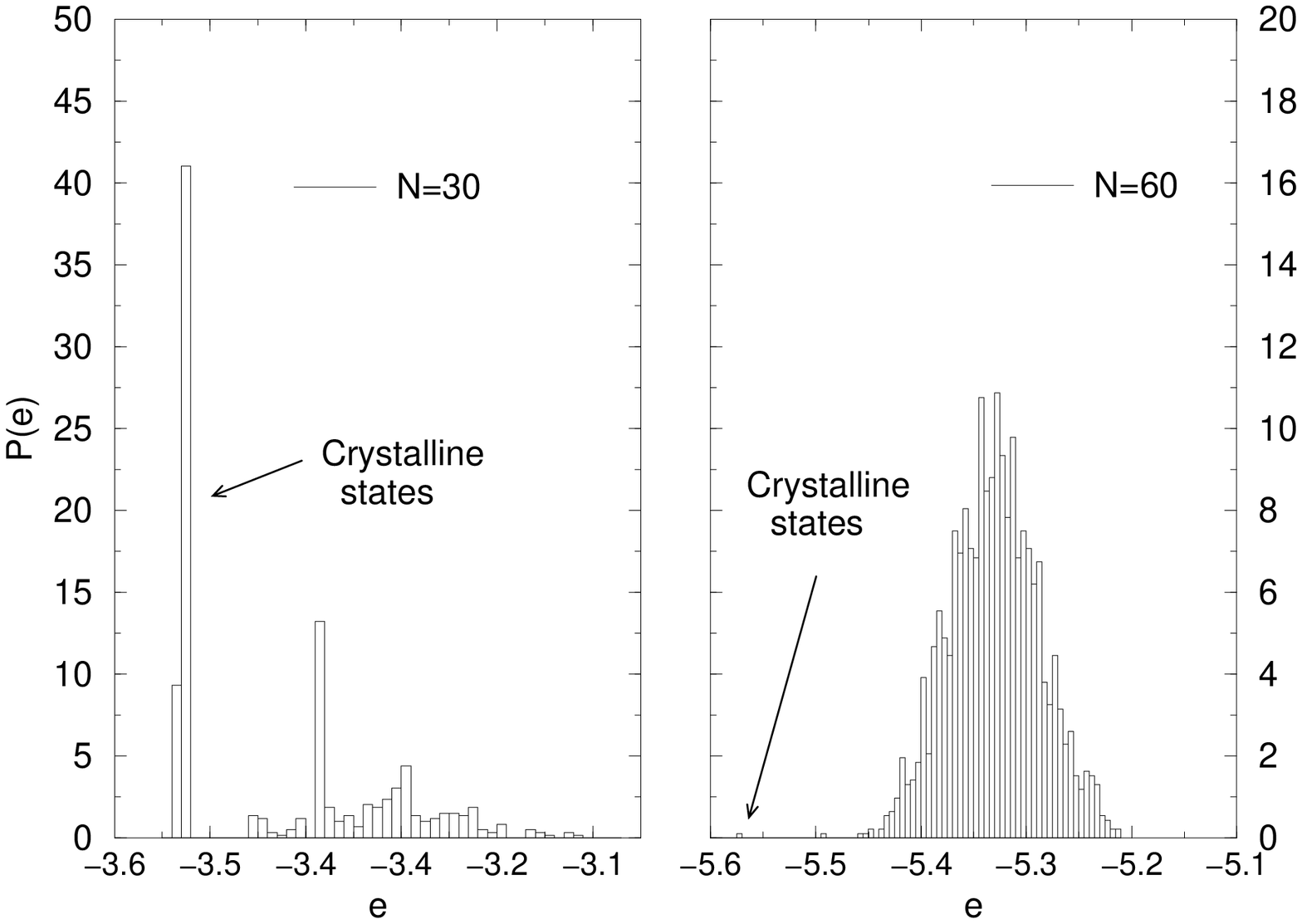}
    \hspace*{4.7mm}
    \epsfysize=5.5cm\epsfbox{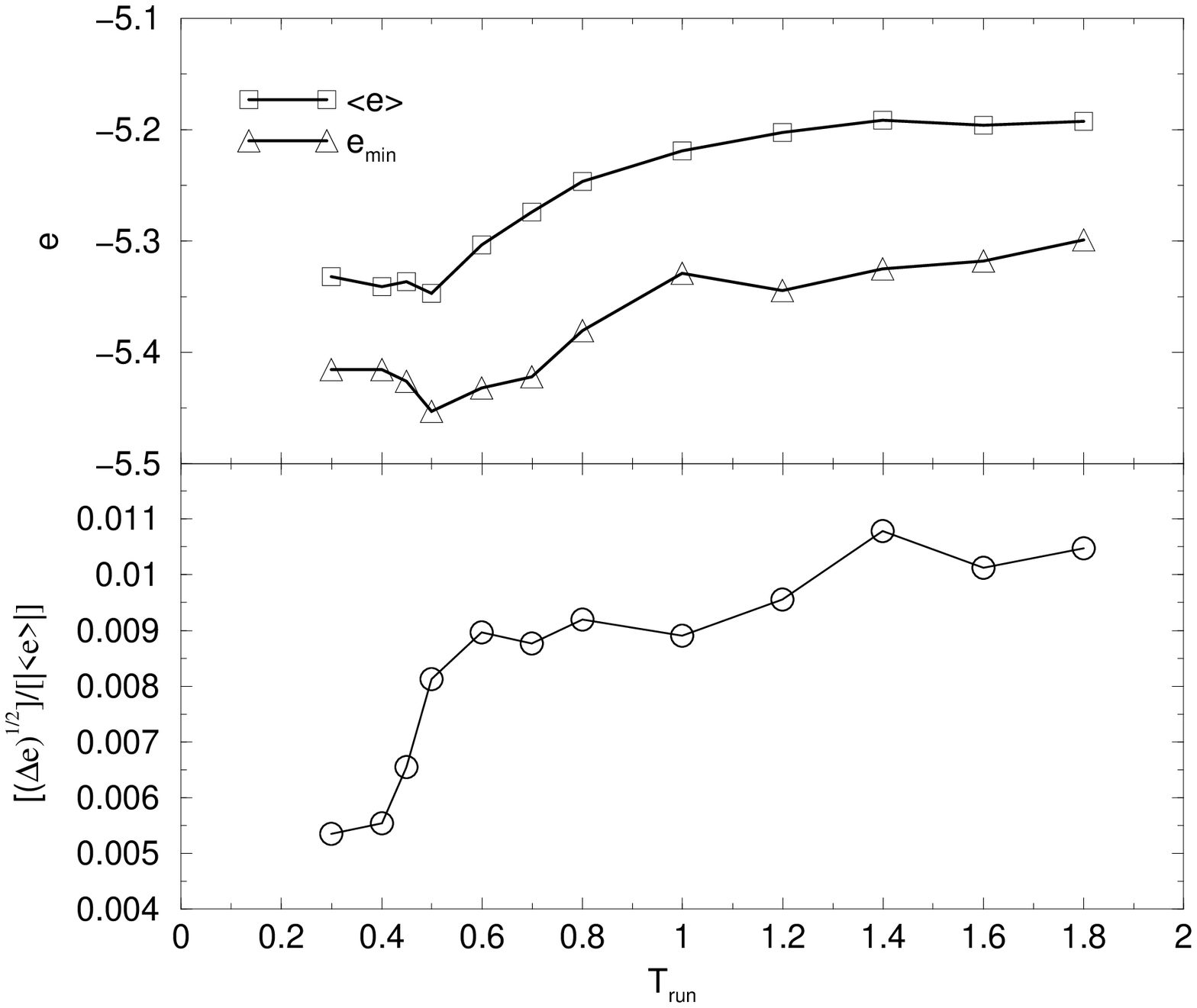}
  \end{center}
  \hbox to \textwidth{\vspace*{5mm}\hfil Fig.\ \ref{energy_spectrum}\hfil\hfil 
  Fig.\ \ref{eges}\hfil}
  \caption{\label{energy_spectrum}
      The energy spectrum of the minima sampled from $T_{\rm run}=0.5$ 
    for $N=30$ (left) and $N=60$ particles (right) is shown as a histogram for
    $1000$ states for each system size.} 
  \caption{\label{eges} 
    Mean energy and variance, as calculated by averaging over $200$
    configurations at each temperature $T_{\rm run}$. The lower graph
    in the upper panel is the lowest energy found in the set of
    configurations at each temperature.}
\end{figure}

In fig. \ref{energy_spectrum} we show a histogram of the energy per particle of
locally minimal states for $N=30$ and $N=60$. For each system size $1000$
minima have been generated, starting from $T_{\rm run}=0.5$. For the small
system, crystalline states have the dominating basins of attraction, whereas
for the larger system we found very few crystalline states (3 out of 1000).

{\it Energy Landscape}.  The set of sampled local minima depends on $T_{\rm
run}$. We have generated $200$ configurations for $N=60$ particles and fixed
$T_{\rm run}$ chosen in the interval $0.3 \le T_{\rm run} \le T=1.8$. For each
$T_{\rm run}$ we have computed the mean energy, the lowest energy found and the
variance of the energy. The results are shown in fig.\ \ref{eges}. For each
configuration the length of the molecular dynamics simulation has been chosen
as $\tau_{\rm run}=0.5*10^6$. This is sufficient to allow for equilibration at
$T_{\rm run}>T_g$.

The mean energies per particle of the sampled minima are approximately
constant ($\langle e \rangle\approx e_h=-5.2$) at high temperatures
and decrease significantly within a transition region. The lowest
energy minima $e_l\approx-5.45$ are found for $T_{\rm run}\approx
T_g$, whereas for $T_{\rm run}<T_g$ sampling is restricted to one
quasi-ergodic component.  Thus extensive sampling of locally stable
states is most effective for $T_{\rm run}\gtrsim T_g$. At high
temperatures, most of the weight is found in a broad band of locally
stable states with high energies around $e_h$ . The low energy
structures are not sampled, although these minima are favoured by a
relative Boltzmann factor $\sim \exp{(N(e_h-e_l)/k_B T_{\rm run})}$.
This implies that the configuration space of the high energy basins is
much larger than that of the low energy basins, so that it
overcompensates the relative Boltzmann factor. When $T_{\rm run}$ is
lowered to temperatures around $T_g$, the Boltzmann factor becomes
more important and eventually overruns the configuration space volumes
of the $e_h$ states. These results are in agreement with recent work
of Sastry {\it et al.} \cite{sds98}, who showed that the transition is
accompanied by the onset of typical, glass like relaxation phenomena.

To assess the performance of our algorithm in sampling low lying energy states
quantitatively, we compare the lowest energies sampled with the results of a
recent study on the optimization properties of traditional simulation
algorithms applied to the Lennard-Jones glasses \cite{bs98}. The authors
reported on a lowest energy configuration found for $N=60$ at $e=-5.38$ by
cooling down slowly from the liquid phase using standard molecular
dynamics. The lowest state shown in fig.\ \ref{eges} is at $e=-5.45$.  The
difference in energy $\delta e = 0.07$ corresponds to a factor of $100$ in
computer time according to the estimate given in refs.\ \cite{bs98,vkb96}. For
$T_{\rm run}\le 0.5$ around $15\%$ of all states sampled using our method are
lower in energy then $e=-5.38$.

{\it Distribution of overlaps}.  To investigate the distribution of overlaps we
have generated $2000$ amorphous states for the $N=60$ particles system keeping
the centre-of-mass fixed. Each molecular dynamics trajectory has been
initialized at $T_{\rm run}=0.5$ and has been run for $\tau_{\rm run}=10^6$
time steps. 
We compute the overlap of the large particles for all pairs $(a,b)$.  This
implies that the weights $P_a$ are approximated by the rate of occurrence of
state $a$ in our simulation.  Several arguments support our choice of ensemble:
1) The generated states are as low in energy as the best ones from other
optimization procedures.  2) In agreement with recent simulations of Sastry
{\it et al.}  \cite{sds98}, we find a qualitative change in the properties of
sampled states at a well defined temperature
$T_g\approx 0.5$, such that above $T_g$ a broad range of states are sampled,
whereas below $T_g$ the system is landscape dominated, {\it i.e.} trapped in
a metastable state.  3) The Stillinger-Weber method is controlled and
reproducible and sufficiently simple to allow for analytical calculations.

\begin{figure}[htbp]
  \begin{center}
    \leavevmode \epsfysize=5cm \epsfbox{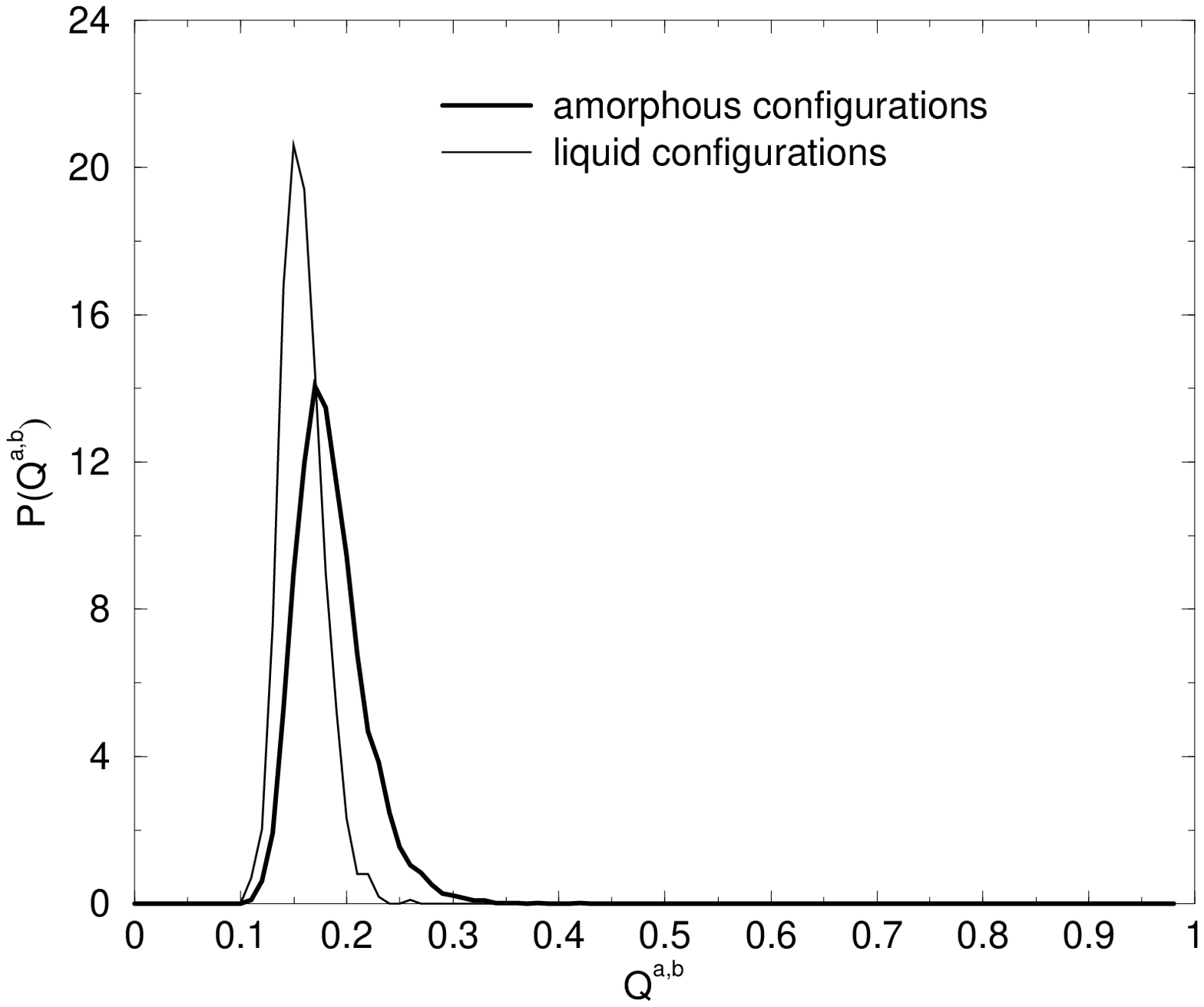}
    \hspace*{1cm}
    \epsfysize=5cm \epsfbox{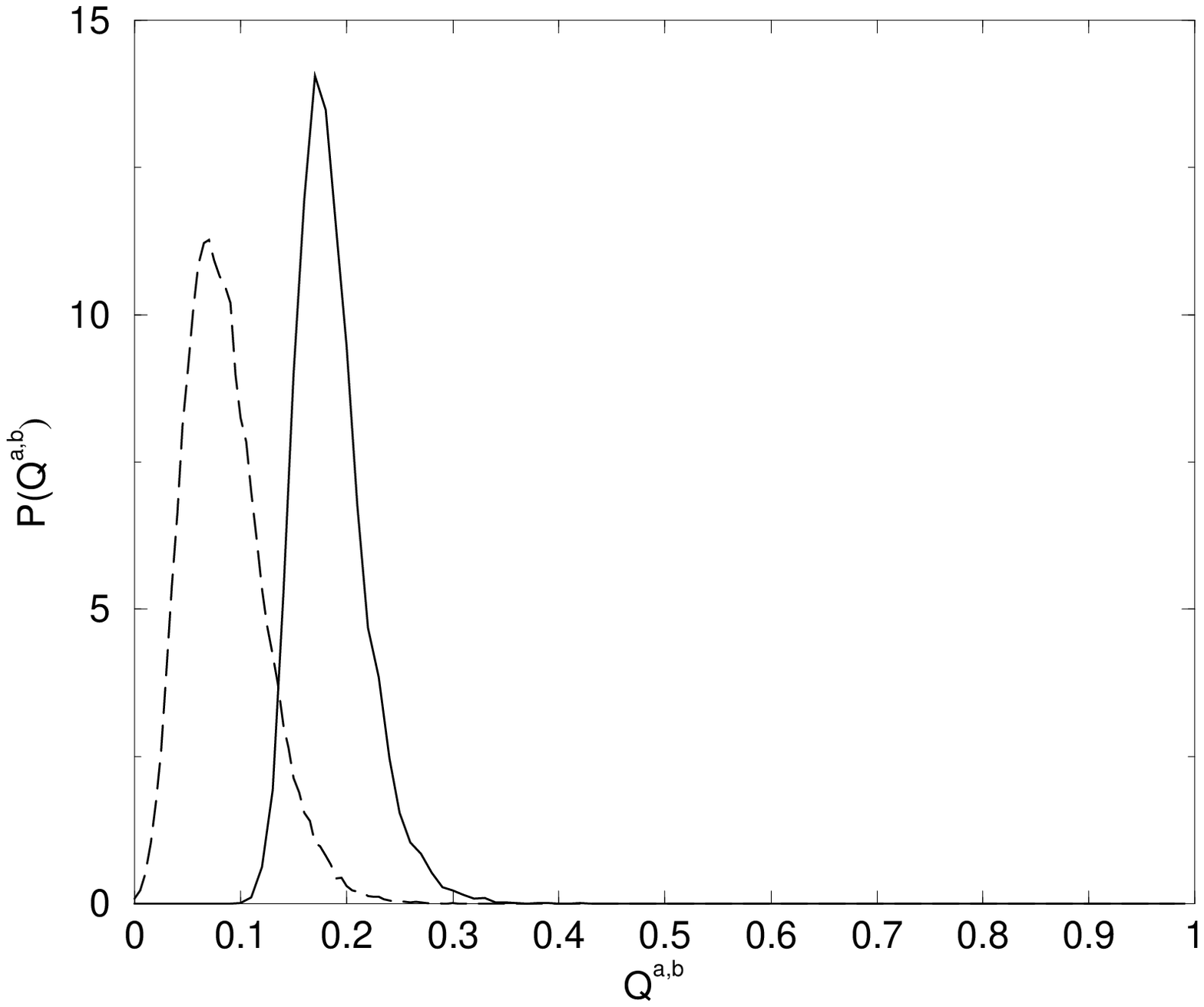}
  \end{center}
 \hbox to \textwidth{\hfil Fig.\ \ref{overlap}\hfil\hfil 
  Fig.\ \ref{rotation}\hfil}
  \caption{\label{overlap}
    The overlap distribution for all glassy states in comparison to the
    overlap distribution of random states, generated as configurations
    of the liquid phase at high temperatures.}
  \caption{\label{rotation}
    The distribution of all overlap computed by either maximizing over the
    $48$ symmetry related states (solid line) or by including all symmetry
    related states with the same weight (dotted line).}
\end{figure}

A histogram of overlaps is shown in fig.\ \ref{overlap}. We observe a most
probable value around
$Q^{a,b}\approx 0.2$ and no significant structure in the distribution. It is
instructive to also compute the distribution of overlaps of liquid
configurations. This distribution is compared to the histogram for glassy
states in fig.\ \ref{overlap}.  The distribution of overlaps for glassy states
is $35\%$ broader and centered at a $20\%$ higher value. Otherwise no
significant changes occur. Thus we conclude that the glassy minima are as
different as they can be. This conclusion also holds, if we restrict our
ensemble of glassy states to the lowest energy states.

The residual cubic point group symmetry does not affect the overlap
distribution significantly. We have calculated overlaps by either maximizing
over the $48$ symmetry related states or by including {\it all} symmetry
related states with the same weight. The resulting distributions are shown in
Fig.\ \ref{rotation}. As one would expect the distribution obtained by
maximizing over all rotations is peaked at a higher value of $Q$, but does not
reveal any additional structure.

\begin{figure}[htbp]
  \begin{center}
    \leavevmode \epsfysize=6cm \epsfbox{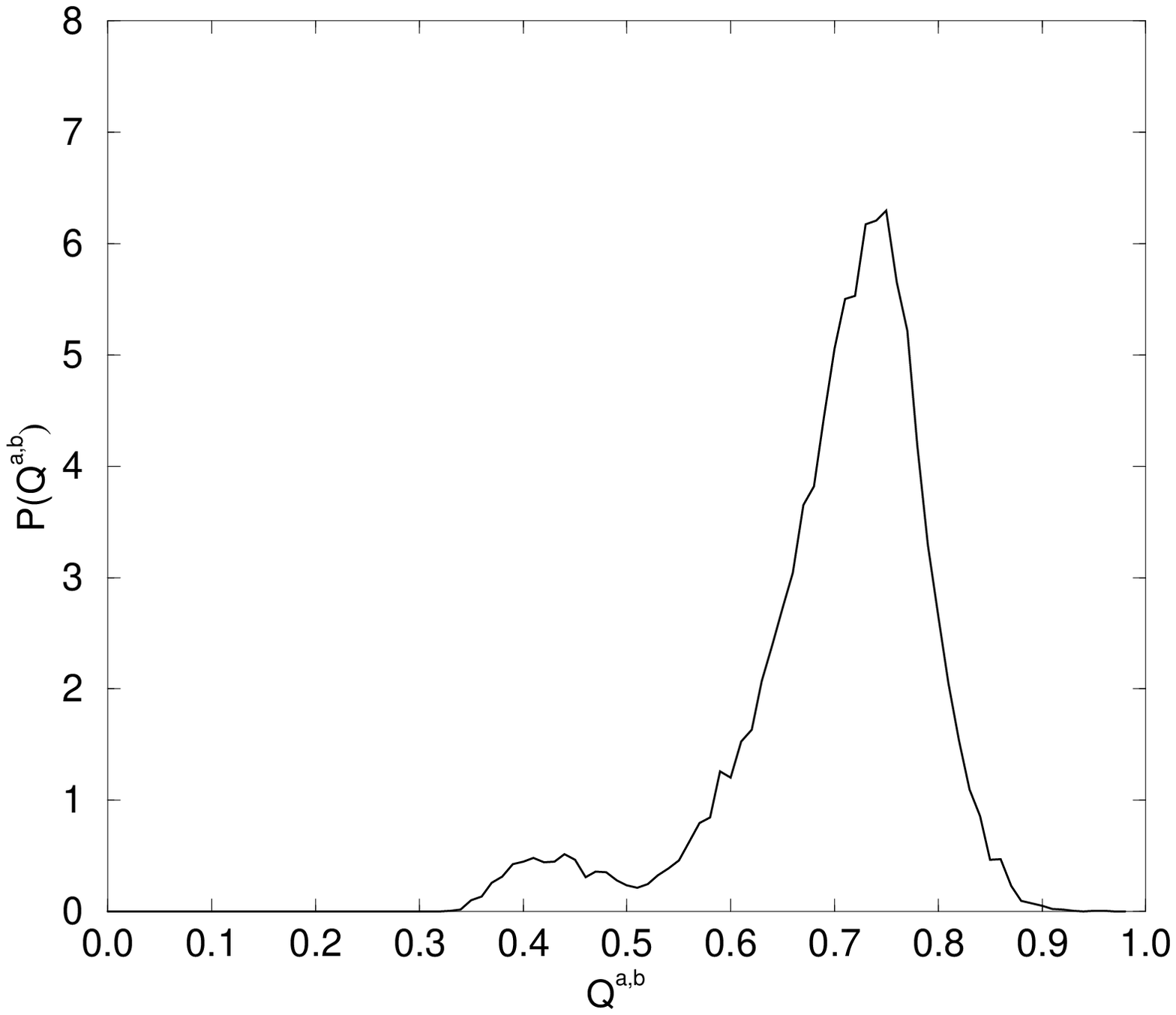}
    \vspace{1cm}
    \caption{\label{Parisi}
      The overlap distribution for a small system of $N=30$ particles,
      confined by soft walls. The dominant peak at high $Q$ is due to
      overlaps between two crystalline states, whereas the lower values of
      $Q$ correspond to overlaps between two amorphous states or one
      amorphous and one crystalline state.}
  \end{center}
\end{figure}

Recently Coluzzi and Parisi \cite{cp98} computed the distribution of
distances $P(D)$, using the distance measure of eq.\ (\ref{mindist})
and simulating small systems ($28$ to $36$ particles) confined by soft
walls. They find highly non-trivial distributions $P(D)$ in the glassy
phase, which strongly depend on particle number. We have also applied
our method to generate an ensemble of low energy states by rapid
cooling for the system of ref.\ \cite{cp98}. A straightforward
evaluation of $P(Q)$ yields the bimodal distribution, shown in fig.\ 
\ref{Parisi}. A closer inspection, however, reveals that the structure
is due to many imperfect crystalline states. Soft walls increase the
probability for crystalline states, as compared to periodic boundary
conditions, and for small systems ($N\approx 30$ particles) most of
the particles sit on the surface of the sample.


\begin{thebibliography}{99}\frenchspacing
\bibitem{kt87} Kirkpatrick T. R. and Thirumalai D., 
  Phys. Rev. Lett., {\bf 58} (1987) 2091; 
  Phys. Rev. B, {\bf 36} (1987) 5388.
\bibitem{p97} Parisi G., cond-mat/9712079 (1997).
\bibitem{sw82} Stillinger F. H. and Weber T. A., 
  Phys. Rev. A, {\bf 25} (1982) 989.
\bibitem{sw84} Stillinger F. H. and Weber T. A., 
  J. Chem. Phys., {\bf 80} (1984) 4434; 
  J. Chem. Phys., {\bf 81} (1989) 5089; 
  Phys. Rev. B, {\bf 31} (1985) 5262.
\bibitem{heuer} Heuer A., 
  Phys. Rev. Lett., {\bf 78} (1997) 4051.
\bibitem{cp98} Coluzzi B. and Parisi G., 
  J. Phys. A: Math. Gen., {\bf 31} (1998) 4349.
\bibitem{at86} Allen M. P. and Tildesley D. J., 
  {\em Computer Simulations of Liquids} 
  (Oxford Science Publications, Oxford) 1996. 
\bibitem{nr} Press W. H., Teukolsky S. A., Vetterling W. T. 
  and Flannery B. P., 
  {\em Numerical Recipes in C: The art of scientific computing}, 
  Second Edition  (Cambridge University Press, Cambridge) 1994. 
\bibitem{ka94} Kob W. and Andersen H. C., 
  Phys. Rev. Lett., {\bf 73} (1994) 13476; 
  Phys. Rev. E, {\bf 51} (1995) 4626; 
  Phys. Rev. E, {\bf 52} (1995) 4134.
\bibitem{vkb96} Vollmayr K., Kob W. and Binder K.,
  J. Chem. Phys., {\bf 105} (1996) 4714.
\bibitem{ha87} Honeycutt J. D. and Andersen H. C., 
  J. Phys. Chem., {\bf 91} (1987) 4915.
\bibitem{sds98} Sastry S., Debenedetti P. G. and Stillinger F. H.,
  Nature, {\bf 393} (1998) 554. 
\bibitem{bs98} Bhattacharya K. K. and Sethna J. P., 
  Phys. Rev. B, {\bf 57} (1998) 2553.
\end{thebibliography}
\end{document}